\documentclass[prb,amsfonts,amssymb,amsmath,eqsecnum,letterpaper]{revtex4}

\usepackage{citesort}
\usepackage{amsmath}

\renewcommand{\r}{\mathbf{r}}

\begin{document}

\title{On the bulk modulus of the cell model of charged macromolecules 
suspensions}

\author{Gabriel T\'ellez}
\email{gtellez@uniandes.edu.co}
\affiliation{Grupo de F\'{\i}sica T\'eorica de la
Materia Condensada, Departamento de F\'{\i}sica, Universidad de Los
Andes, A.A. 4976, Bogot\'a, Colombia}

\author{Emmanuel Trizac}
\email{Emmanuel.Trizac@th.u-psud.fr}
\affiliation{
Laboratoire de Physique Th\'eorique, B\^atiment 210, 
Universit\'e de Paris-Sud, 91405 Orsay, France (Unit\'e Mixte de 
Recherche 8627 du CNRS)}


\begin{abstract}
We study theoretically the bulk modulus (inverse of the
compressibility) of a suspension of charged objects (macro-ions),
making use of a cell model to account for the finite density of
macro-ions.  The diffuse layer of charged micro-species around a
macro-ion is described by a generic local density functional
theory. Within this general framework, we obtain the condition for a
positive bulk modulus, which is fulfilled by several proposals made in
the literature and rules out the possibility of a critical point.  We
show that a sufficient condition for a positive compressibility also
ensures that the same theory produces repulsive effective pair
potentials.
\end{abstract}

\maketitle

%%%%%%%%%%%%%%%%%%%%%%%%%%%%%%%%%%%%%%%%%%%%%%%%%%%%%%%%%%%%%%%%%%%%%%%%%%%%%%%
\section{Introduction}
Macromolecules soluble in aqueous suspensions usually acquire 
an electric charge. Such systems are of considerable theoretical and 
experimental interest. Examples include proteins in living cells, 
dispersion paints or superabsorbants. Their theoretical description 
is however a tremendous task, and operational approximate 
treatments are needed. 

In this work, we consider a cell model to account for the macro-ions
correlations \cite{Marcus}, in conjunction with a local density functional
theory to describe the inhomogeneous electric double-layer around 
a macro-ion \cite{Stevens,Trizac-forces}. This framework encompasses the 
standard mean-field Poisson-Boltzmann (PB) theory, but also more 
recent approaches proposed to account for micro-ions 
excluded volume \cite{Eigen}, more general 
non electrostatic effects \cite{Lue}
or to incorporate correlations between the screening micro-ions 
\cite{Barbosa,Barbosa2}, that are neglected within PB theory.
The cell model description may 
be considered as one of
the simplest starting points and provides an important benchmark against 
which experiments and more refined theories are tested. The resulting 
differential equations are however highly non-linear, 
and even with the simplest
of the approaches under consideration here (PB), 
can only be solved analytically
in 1 or 2 dimensions without added salt ({\it i.e.} for the situation of 
a flat double-layer in a confining slab, or for that of a rod-like
macro-ion of infinite length enclosed in a concentric confining 
cylinder \cite{Katchalsky}).
The linearized version of the above problem has therefore always been an important
alternative, but is not free of internal inconsistencies. 
In particular, the linearized PB theory may lead to negative osmotic pressures
\cite{Trizacbis,Deserno} and negative bulk modulus (inverse of the 
compressibility) \cite{Deserno}, whereas within the original non-linear PB
theory, it is straightforward to show that the osmotic pressure is 
necessarily positive. For the bulk modulus, the situation is less 
clear: although  this quantity is found positive in the numerical solution
of PB equation, it seems that 
no formal proof exists concerning its sign. In this article, we derive 
such a proof and obtain the conditions under which a general 
local density functional
theory leads to a positive compressibility within the cell model. 

The article is organized as follows. The density functional theory 
formalism is presented in section \ref{sec:dft} where a few useful 
identities are derived. The bulk modulus is then computed in 
section \ref{sec:comp} and cast in a form where a sufficient
condition for its positivity clearly appears. Conclusions are drawned 
in the the final section.

%%%%%%%%%%%%%%%%%%%%%%%%%%%%%%%%%%%%%%%%%%%%%%%%%%%%%%%%%%%%%%%%%%%%%%%%%%
\section{General formalism}
\label{sec:dft}
\subsection{Density functional theory}

We consider a $\nu$-dimensional colloid with spherical symmetry confined
in its (concentric) Wigner-Seitz cell ($\nu=2$ for a cylindrical colloid, 
$\nu=3$ for a colloidal sphere). The cell is a spherical region ${\cal R}$ of
radius $R$ and volume $V=S_{\nu} R^{\nu}/\nu$ ($S_{\nu}$ is the area
of the unit radius $\nu$-dimensional sphere, $S_{2}=2\pi$,
$S_{3}=4\pi$). The colloid (of fixed uniform surface charge density) 
is immersed in an electrolyte solution with
several different species of ions with charges $\{q_{\alpha}\}$, and the
local density of the species $\alpha$ is denoted $n_{\alpha}(\r)$. The system
inside the cell is globally neutral. In what follows, we explicitly 
consider the semi grand-canonical situation where the macro-ions suspension
is in osmotic equilibrium with a salt reservoir through a semi-permeable 
membrane (permeable to micro-species only). We therefore consider the 
grand potential $\Omega\left\{n_{\alpha}\right\}$, which is a functional
of micro-ion densities that we write
\begin{equation}
\label{eq:gd-pot}
\Omega\left\{n_{\alpha}\right\}=
\int_{\cal R} \omega(\{n_{\alpha}(\r)\}) \, d\r +
\frac{1}{2} \int_{{\cal R}} \rho(\r) \psi(\r) \,d\r
+ \lambda\int_{{\cal R}} \rho(\r)\,d\r,
\end{equation}
where
\begin{equation}
\rho(\r)=\sum_{\alpha} q_{\alpha} n_{\alpha}(\r) +
\rho_{\mathrm{col}}(\r)
\end{equation}
is the total charge density including that of the colloid (denoted
$\rho_{\mathrm{col}}$), $\lambda$ is a Lagrange multiplier to ensure
global electroneutrality ($\partial\Omega/\partial \lambda = \int \rho
=0$) and $\psi(\r)$ is the electrostatic potential. Within PB theory,
the local part of the functional (\ref{eq:gd-pot}) embodied in
$\omega$ is entirely of entropic origin. In other words, electrostatic
interactions are taken into account at mean-field level only, through
the term $\int \rho \psi/2$. This feature is shared by the theories of
Refs.~[\onlinecite{Eigen,Lue}], but is not the case of the formalism put
forward by Barbosa {\it et al.} \cite{Barbosa,Barbosa2}, 
where the term $\omega$ contains
electrostatic corrections to the mean-field Coulomb contribution. By
comparison with the results of Molecular Dynamics simulations, this
framework was shown to capture important correlations missed by the
mean-field Poisson-Boltzmann \cite{Barbosa}.

Within the standard Poisson-Boltzmann theory, the micro-ions are
treated as an ideal gas of charged particles reacting to the mean
electrostatic potential. The free energy density therefore reads
$f(\{n_{\alpha}\}) = f_{\mathrm{id}}(\{n_{\alpha}\})=
\beta^{-1}\sum_{\alpha} n_{\alpha} \{\ln [\Lambda^3 n_{\alpha}]-1\}$
where $\Lambda$ is an irrelevant length scale and $\beta=1/(k_B T)$
the inverse temperature.  The grand potential functional describing
the osmotic equilibrium with a salt reservoir is thus $\Omega = \int f
-\sum_{\alpha} \mu_\alpha^b \int n_{\alpha} + \int \rho \psi/2
+\lambda \int \rho$, where the chemical potentials $\mu_\alpha^b =
\partial f_{\mathrm{id}}(\{n_{\alpha}^b\})/\partial n_{\alpha}^{b} =
\beta^{-1} \ln (\Lambda^3 n_{\alpha}^b)$ are defined from the bulk
densities $n_\alpha^b$ of micro-species in the reservoir.  The local
grand potential density finally takes the form
\begin{equation}
\omega(\{n_{\alpha}(\r)\})= f(\{n_{\alpha}(\r)\}) - 
\sum_{\alpha} \mu_\alpha^b n_{\alpha}(\r) = 
\beta^{-1}\sum_{\alpha} n_{\alpha}(\r)
\left(\ln\frac{n_{\alpha}(\r)}{n_{\alpha}^b}-1\right).
\end{equation}
As mentioned in the introduction, other theories may be described
by our formalism, such as those proposed to account for
steric effects~\cite{Eigen}, more general non-electrostatic
interactions~\cite{Lue} or to go beyond mean field and include
correlations~\cite{Barbosa,Barbosa2}.

The mean-field Coulomb term $\frac{1}{2}\int \rho(\r)\psi(\r)\,d\r$ in
Eq.  (\ref{eq:gd-pot}) is actually quadratic in the densities if one
writes the potential in term of the set $\{n_\alpha\}$. Introducing
the proper Green's function $G(\r,\r')$ for the electrostatic problem
in the region ${\cal R}$, it is always possible to write the
electrostatic potential in the form (see appendix \ref{app:a})
\begin{equation}
\psi(\r)=\int_{\cal R} \rho(\r')\, G(\r,\r')\,d\r',
\label{eq:convol}
\end{equation}
with the cell boundary chosen as the potential reference, $\psi(R)=0$.

The variational problem defined by the functional~(\ref{eq:gd-pot})
leads to the stationary condition
\begin{equation}
\label{eq:minimiz}
\frac{\partial\omega(\{n_{\alpha}(\r)\})}{\partial n_{\alpha}}
+q_{\alpha} (\psi(\r) + \lambda)=0,
\end{equation}
which, in the Poisson-Boltzmann theory, reduces to 
\begin{equation}
n_{\alpha}(\r)=
n_{\alpha}^b e^{-\beta q_{\alpha}[ \psi(\r)+\lambda]}.
\end{equation}
Since the potential vanishes at the boundary of the cell, it can be
seen in this equation that the Lagrange multiplier $\lambda$ coincides
with the so-called Donnan potential (potential drop across the
interface of the cell and the bulk reservoir, which is formally at a
potential $-\lambda$).

Finally, discriminating between the optimal density profiles fulfilling
Eq.~(\ref{eq:minimiz}) and the generic arguments of the grand potential
functional leads to unnecessary heavy notations and will not be
useful for the subsequent analysis.

%%%%%%%%%%%%%%%%%%%%%%%%%%%%%%%%%%%%%%%%%%%%%%
\subsection{A few useful identities}

Consider a functional ${\cal Q}$ of the density profiles
$n_{\alpha}(\r)$ having an explicit dependence on the volume $V$ of
the cell and the potential drop $\lambda$ (later, we shall be 
interested in ${\cal Q}=\Omega$). The total derivative $d {\cal Q}/d V$ of
${\cal Q}$ when the volume of the cell varies is~\cite{Deserno}
\begin{equation}
\label{eq:variation-V}
\frac{d{\cal Q}}{d V}=
\frac{\partial {\cal Q}}{\partial V} 
+
\int_{{\cal R}}
\sum_{\alpha} \frac{\delta {\cal Q}}{\delta n_{\alpha}(\r)} 
\frac{d n_{\alpha}(\r)}{d V}\,d\r
+
\frac{\partial {\cal Q}}{\partial \lambda} 
\frac{d\lambda}{d V}
\end{equation}
where the first term is the variation due to the explicit dependence
of ${\cal Q}$ on $V$, the second is due to the variation of the
density profiles when the volume changes, and the last
is due to the variation of the potential drop with the
volume. Computing $d n_{\alpha}(\r)/d V$ and $d\lambda/dV$ would
require to solve the variational problem~(\ref{eq:minimiz}) subject to
the neutrality condition for a cell of volume $V+dV$ and for a cell of
volume $V$, before computing the difference of 
the two solutions. However this
will not be necessary for our purposes.

We will be interested most of the time in quantities defined by a
local density. In general the partial derivative with respect to the
volume (explicit dependence) of such a quantity may be computed by
means of the dilatation method:
\begin{eqnarray}
\frac{\partial}{\partial V}
\left[
\int_{{\cal R}} g(\r)\,d\r
\right]
&=&
\frac{\partial}{\partial V}
\left[ V
\int_{\tilde{{\cal R}}} g(V^{1/\nu}\tilde{\r})\,d\tilde{\r}
\right]
\nonumber\\
&=&
\frac{1}{V} \int_{{\cal R}} g(\r)\,d\r +
\frac{1}{V\nu}
\int_{{\cal{R}}} \r\cdot\frac{\partial g(\r)}{\partial \r}\,d\r
\nonumber\\
&=&
\frac{1}{V} \int_{{\cal R}} g(\r)\,d\r +
\frac{1}{V\nu} \int_{\partial{\cal R}} g(\r) \r\cdot d\mathbf{S}
-\frac{1}{V\nu} \int_{{\cal R}}  \nu g(\r)\,d\r \\
\nonumber\\
&=&
\frac{1}{V\nu} \int_{\partial{\cal R}} g(\r) \r\cdot d\mathbf{S}.
\end{eqnarray}
We first made a change of variable $\r=V^{1/\nu}\tilde{\r}$ to show
explicitly the volume dependence of the integral. After computing the
derivatives and returning to the unscaled variable $\r$ we made an
integration by parts ($\nu$ is the dimensionality of the cell $\nu= 2,
3$). For the spherical isotropic geometry we are interested in, this
result reduces to
\begin{equation}
\label{eq:derivee-V-local}
\frac{\partial}{\partial V}
\left[
\int_{{\cal R}} g(\r)\,d\r
\right]
=
g(R).
\end{equation}
We will need also to compute the derivative of some terms given by a
double integral. Following the same steps as above, 
this derivative is given by
\begin{equation}
\label{eq:derivee-V-bilocal}
\frac{\partial}{\partial V}
\left[
\int_{{\cal R}} g(\r,\r')\,d\r\,d\r'
\right]
=
\int_{\cal R} g(R,\r')\,d\r'
+
\int_{\cal R} g(\r,R)\,d\r. 
\end{equation}
The key assumption here is that both quantities $\int g(\r,\mathbf{R})
\,d\r$ and $\int g(\mathbf{R},\r) \,d\r$ only depend on the modulus
$R=|\mathbf{R}|$.  Another important point to notice is that the
volume total derivative is taken at constant total electric charge
because the system is neutral. Applying
equation~(\ref{eq:variation-V}) to $\sum_{\alpha}
q_{\alpha}N_{\alpha}=\int \sum_{\alpha} q_{\alpha}
n_{\alpha}(\r)\,d\r$ then gives
\begin{equation}
\label{eq:variation-N}
\sum_{\alpha} q_{\alpha} \frac{d N_{\alpha}}{d V} =0=
\rho(R)+\int_{\cal R} \sum_{\alpha} q_{\alpha} \frac{d
n_{\alpha}(\r)}{d V} \,d\r.
\end{equation}

%%%%%%%%%%%%%%%%%%%%%%%%%%%%%%%%%%%%%%%%%%%%%%%
\section{The compressibility}
\label{sec:comp}

\subsection{Derivation from the grand potential}

Applying formula~(\ref{eq:variation-V}) to the grand potential once
gives minus the pressure. However, as noticed in
Ref.~[\onlinecite{Deserno}], the second and third terms vanish at the
solution of the variational problem. Making use of
equations~(\ref{eq:derivee-V-local}) and~(\ref{eq:derivee-V-bilocal}),
we consequently have
\begin{subequations}
\begin{eqnarray}
\frac{\partial\Omega}{\partial V} &=& \omega(\left\{n_{\alpha}(R)\right\}) +
\rho(R)[\psi(R)+\lambda]
\\
\label{eq:derivee-premiere}
&=& \omega(\left\{n_{\alpha}(R)\right\}) +
\lambda\rho(R).
\end{eqnarray}
\end{subequations}
The last equality follows from $\psi(R)=0$ 
[see the property~(\ref{eq:propriete-utile})]
and is valid in general for arbitrary isotropic densities
\textit{even} if they differ from the optimal ones solving 
the variational problem. This will be used later. Note that
the symmetry property of the Green's function [$G(\r,\r') = G(\r',\r)$]
is an important ingredient in obtaining the above equations
\cite{Kim}.

Using the stationary condition~(\ref{eq:minimiz}) one finds the usual
result~\cite{Marcus,Deserno,Trizac-forces}
\begin{equation}
p=\sum_{\alpha} n_{\alpha}(R) 
\frac{\partial \omega}{\partial n_{\alpha}(R)}
-\omega(\left\{n_{\alpha}(R)\right\})
\end{equation}
which reduces  to
$
p= k_B T \sum_{\alpha} n_{\alpha} (R)
$ 
within PB theory (see Ref.~[\onlinecite{Linse}] for a general derivation 
of this result, valid beyond PB). 

When computing the second total derivative with respect to the volume
of $\Omega$ to obtain the bulk modulus, one should not disregard 
the second and third terms of
equation~(\ref{eq:variation-V}) to early.
\begin{subequations}
\begin{eqnarray}
\frac{d^2 \Omega}{d V^2}
&=&
\label{eq:terme-1}
\frac{\partial^2 \Omega}{\partial V^2}
\\
&&
\label{eq:terme-2-4}
+2
\int_{\cal R} \sum_{\alpha}
\frac{\delta}{\delta n_{\alpha}(\r)}\left[
\frac{\partial \Omega}{\partial V}\right]
\frac{d n_{\alpha}(\r)}{dV}\, d\r
\\&&
\label{eq:terme-3-6}
+2 \frac{\partial^2 \Omega}{\partial V\partial \lambda}
\frac{d\lambda}{dV}
\\&&
\label{eq:terme-0}
+
\int_{{\cal R}^2}
\sum_{\alpha\gamma}
\frac{\delta^2 \Omega}{\delta n_{\alpha}(\r) \delta n_{\gamma}(\r')}
\frac{d n_{\alpha}(\r)}{dV} \frac{d n_{\gamma}(\r')}{dV}
\,
d\r\,d\r'
\\&&
\label{eq:terme-5-7}
+2
\int_{\cal R} \sum_{\alpha}
\frac{\delta}{\delta n_{\alpha}(\r)}\left[
\frac{\partial \Omega}{\partial \lambda}
\right]
\frac{d n_{\alpha}(\r)}{dV}
\frac{d\lambda}{dV}
\, d\r
\\&&
\label{eq:terme-8}
+
\frac{\partial^2\Omega}{\partial\lambda^2}
\left[\frac{d\lambda}{dV}\right]^2.
\end{eqnarray}
\end{subequations}
At the solution of the variational problem most of these terms vanish,
as it will be shown below.  The first term~(\ref{eq:terme-1}) is
obtained taking the partial derivative of~(\ref{eq:derivee-premiere})
with respect to $V$
\begin{equation}
\frac{\partial}{\partial V}\left[\omega(\{n_{\alpha}(R)\})
+
\lambda\rho(R)\right]
=
\frac{1}{S_{\nu} R^{\nu-1}}
\left\{
\sum_{\alpha} \frac{\partial n_{\alpha}(R)}{\partial R}
\left[\frac{\partial\omega}{\partial n_{\alpha}(R)}
+q_{\alpha}
\lambda\right]
\right\}
=0.
\end{equation}
We have used the variational equation~(\ref{eq:minimiz}) at
$\r=\mathbf{R}\in\partial{\cal R}$ and the fact that $\psi(R)=0$. The
second term~(\ref{eq:terme-2-4}) is obtained
replacing the expression~(\ref{eq:derivee-premiere}) for
$\partial\Omega/\partial V$ into~(\ref{eq:terme-2-4})
\begin{equation}
2\sum_{\alpha}
\left[
\frac{\partial\omega}{\partial n_{\alpha}(R)}
+
q_{\alpha}
\lambda
\right]
\frac{d n_{\alpha}(R)}{dV}
=
0
\end{equation}
where we have once more used equation~(\ref{eq:minimiz}) at
$\r=\mathbf{R}\in\partial{\cal R}$.

The third term~(\ref{eq:terme-3-6}) is equal to 
\begin{equation}
2\rho(R) \frac{d\lambda}{dV}
\end{equation}
and with the fifth term~(\ref{eq:terme-5-7})
\begin{equation}
2\frac{\partial}{\partial \lambda}
\int_{\cal R}
\sum_{\alpha}
\left[
\frac{\partial\omega}{\partial n_{\alpha}}
+ q_{\alpha}\left(\psi(\r)+\lambda\right)
\right]
\frac{d n_{\alpha}(\r)}{dV}\,
\frac{d\lambda}{dV}
\,d\r
=
2\int_{\cal R}
\sum_{\alpha}
q_{\alpha}\frac{d n_{\alpha}(\r)}{dV}
\frac{d\lambda}{dV}
\,d\r
=-2\rho(R)
\frac{d\lambda}{dV}
\end{equation}
gives a vanishing contribution. We have used in the preceding equation the
relation~(\ref{eq:variation-N}).
The last term~(\ref{eq:terme-8}) is zero since $\Omega$ in linear on
$\lambda$.
Finally the inverse compressibility may be cast in the form
\begin{equation}
\label{eq:compress-1}
\chi^{-1}
=
-V\left(\frac{\partial p}{\partial V}\right)_{T,\mu_i}
=
V\frac{d^2\Omega}{dV^2}
=
V\int_{{\cal R}^2}
\sum_{\alpha\gamma}
\frac{\delta^2 \Omega}{\delta n_{\alpha}(\r) \delta n_{\gamma}(\r')}
\frac{d n_{\alpha}(\r)}{dV} \frac{d n_{\gamma}(\r')}{dV}
\,
d\r\,d\r'.
\end{equation}
{}From equation~(\ref{eq:gd-pot}) we have
\begin{equation}
\frac{\delta^2 \Omega}{\delta n_{\alpha}(\r) \delta n_{\gamma}(\r')}
=
\frac{\partial \omega(\{n_{\delta}(\r)\})}{\partial n_{\alpha}\partial
n_{\gamma}}\,\delta(\r-\r')
+
q_{\alpha} q_{\beta} G(\r,\r')
\end{equation}
The second term, the Coulomb contribution, when replaced into
equation~(\ref{eq:compress-1}) gives
\begin{subequations}
\begin{eqnarray}
\label{eq:Coulomb-contrib-a}
\int_{{\cal R}^2}
\sum_{\alpha\gamma}
q_{\alpha}q_{\gamma}
G(\r,\r')
\frac{d n_{\alpha}(\r)}{dV}
\frac{d n_{\gamma}(\r')}{dV}
\,d\r\,d\r'
&=&
-
\frac{1}{S_{\nu}}
\int_{\cal R}
\phi(\r)\Delta \phi(\r)\,d\r
\\
\nonumber
&=&
-R^{\nu-1}\phi(R)\partial_n\phi(R)
+
\frac{1}{S_{\nu}}
\int_{\cal R} \left|\nabla \phi(\r)\right|^2
\,d\r
\\
\label{eq:Coulomb-contrib-b}
\end{eqnarray}
\end{subequations}
where we have defined the ``electric potential'' created by the charge
variation
\begin{equation}
\phi(\r)=\int_{\cal R} G(\r,\r')
\sum_{\alpha}
q_{\alpha}
\frac{d n_{\alpha}(\r')}{dV}
\, d\r'
\end{equation}
and performed an integration by parts.  The boundary term
in~(\ref{eq:Coulomb-contrib-b}) vanishes because, from
equation~(\ref{eq:propriete-utile}), one has
\begin{equation}
\phi(R)= \int \sum_\alpha
q_\alpha\frac{dn_{\alpha}(\r')}{dV}\,G(\mathbf{R},\r')\, d\r'=0,
\end{equation}
thus showing that the Coulomb contribution term is always positive and
\begin{equation}
\label{eq:compress-2}
\chi^{-1}
=
V
\int_{\cal R}
\sum_{\alpha\gamma}
\frac{\partial^2\omega}{\partial n_{\alpha}\partial n_{\gamma}}
\frac{d n_{\alpha}(\r)}{dV}
\frac{d n_{\gamma}(\r)}{dV}
+
\frac{V}{S_{\nu}}
\int_{\cal R}
\left|
\nabla_{\r}
\left[
\int_{\cal R}
G(\r,\r') \sum_{\alpha}
q_{\alpha} 
\frac{d n_{\alpha}(\r')}{dV}
\,d\r'
\right]
\right|^2 \,d\r.
\end{equation}

%%%%%%%%%%%%%%%%%%%%%%%%%%%%%%%%%%%%%%%%%%%%%
\subsection{Discussion}
From Eq. (\ref{eq:compress-2}), the positive definiteness 
of the integral operator whose kernel is defined by 
\begin{equation}
\frac{\partial^2\omega(\{n_{\delta}(\r)\})}{\partial n_{\alpha}\partial
n_{\gamma}} \delta(\r-\r')
\end{equation}
ensures that the compressibility is positive (this
is a sufficient but not necessary condition). This is the case for 
Poisson-Boltzmann theory where $\omega$ is simply the ideal
gas grand potential density, and it may be checked that it also
holds for the theories
presented in Refs.~[\onlinecite{Eigen,Lue,Barbosa}].

More generally, in any well constructed approximate theory for the
colloid based on density functionals for the grand potential of the
form~(\ref{eq:gd-pot}) (that is a local density term plus an
``interaction term'' given by the mean-field 
Coulomb electrostatic energy), the
solution of the variational problem should be a minimum, i.~e.~the
quadratic form
\begin{equation}
\frac{\delta^2 \Omega}{\delta n_{\alpha}(\r) \delta n_{\gamma}(\r')}
\end{equation}
should be positive definite to ensure thermodynamic stability. From
expression~(\ref{eq:compress-1}), it then follows that the
compressibility will be always positive in such a theory.

Equation~(\ref{eq:compress-1}) for the compressibility can be seen as
a generalization for non-uniform fluids of the 
compressibility sum rule for uniform fluids \cite{Henderson}
\begin{equation}
\beta (n\chi)^{-1}=1-n\int c^{(2)}(r)d\r
\end{equation}
written here for a one-component system. 
In this relation, $c^{(2)}(r)$ is the direct
correlation function defined, in the more general situation of 
a mixture, from
\begin{equation}
\label{eq:def-direct-correl}
\frac{\delta^2\Omega}{\delta n_{\alpha}(\r)\delta n_{\gamma}(\r')}
=
\frac{\delta^2{\cal F}_{\mathrm{id}}}{\delta 
n_{\alpha}(\r)\delta n_{\gamma}(\r')}
-k_B T c_{\alpha \gamma}^{(2)}(\r,\r')
\end{equation}
where ${\cal F}_{\mathrm{id}}$ is the ideal gas contribution to the
free energy functional.
When one replaces~(\ref{eq:def-direct-correl})
into~(\ref{eq:compress-1}) for a uniform fluid
[$c^{(2)}_{\alpha\gamma}(\r,\r')=c^{(2)}_{\alpha\gamma}(|\r-\r'|)$],
and knowing that for a uniform fluid
$dn_{\alpha}(\r)/dV=-N_{\alpha}/V^2=-n_{\alpha}/V$, one recovers the
compressibility sum rule
\begin{equation}
\beta \chi^{-1}=
\sum_{\alpha} n_{\alpha}
-
\int_{\cal R}
\sum_{\alpha\gamma}
n_{\alpha} n_{\gamma}
c^{(2)}_{\alpha\gamma}(r)\, d\r.
\end{equation}
Although equation~(\ref{eq:compress-1}) is a natural generalization of
the compressibility sum rule and similar expressions exist in the
literature (see for instance Ref.~[\onlinecite{Henderson}]) we included
here the derivation of this result in the context of the
Poisson-Boltzmann and other generic local density functional theories
because these theories are approximate (non-exact) and nothing
garantees in advance the validity of equation~(\ref{eq:compress-1})
for non-exact theories.

\section{Conclusion}

Within the cell model and a generic local density functional theory, 
we have considered a suspension of charged spherical macro-molecules 
of arbitrary dimension. We have cast the corresponding 
compressibility in a form where the sign of this quantity 
is positive under a (weak) sufficient condition: the grand 
potential $\omega$ appearing in Eq. (\ref{eq:gd-pot}) should
be a convex-up function on densities $\{n_\alpha\}$.
This proves
that the stability requirement of a positive compressibility
is fulfilled by Poisson-Boltzmann theory as well as several 
improvements upon this mean-field approach 
\cite{Eigen,Lue,Barbosa,Barbosa2}.
Our results show that such theories yield stable suspensions,
and cannot exhibit a critical point associated with a gas-liquid
phase separation.

This result should be put in perspective with the recent proofs that
within PB theory, the effective interactions between two identical
colloids confined in a cylinder of infinite length is necessarily
positive \cite{Neu,Sader}. This proof has been extended to the more
general family of local density functional theories (\ref{eq:gd-pot})
in Refs.~[\onlinecite{Trizac-forces,Lang}]. It was shown that the
positive definiteness of the local free energy density (or
equivalently of the grand potential density in the semi-grand
canonical situation) was a sufficient condition for repulsive
interactions. The results derived in this article show that under the
same circumstances exactly, the bulk modulus within a cell model is
also a positive quantity.

We explicitly considered the semi grand-canonical situation where the
macro-ions suspension is dialyzed against an electrolyte
reservoir. Our results may however be extended to other electrostatic
situations, such as the canonical one where the mean salt content in
the suspension is fixed, or to the situation where the macro-ions are
held at constant potential rather than constant charge.

%%%%%%%%%%%%%%%%%%%%%%%%%%%%%%%%%%%%%%%%%%%%%%%%%%%%%%%%%%%%%%%%%%%%%%%%
\begin{appendix}
\section{On the choice of the Green's function}
\label{app:a}
We consider the Green's function $G(\r,\r')$ satisfying 
\begin{equation}
\nabla^2_\r G(\r,\r') = - S_\nu \,\delta(\r-\r')
\end{equation}
with yet unspecified boundary conditions.  From a standard identity
(Green second identity), the solution of Poisson's equation
$\nabla^2_\r \psi = - S_\nu \rho(\r)$ obeys the
relation~\cite{Jackson-book}
\begin{equation}
\psi(\r)= \int_{\cal R} \rho(\r')
\,G(\r,\r')\,d\r' +\frac{1}{S_{\nu}} \int_{\partial{\cal R}}
\frac{\partial \psi(\r')}{\partial n'}\,G(\r,\r')\,dS'
-\frac{1}{S_{\nu}} \int_{\partial{\cal R}} \psi(\r')\,\frac{\partial
G}{\partial n'}(\r,\r')\,dS'
\end{equation}
where $\partial_n \equiv \partial/\partial n$ denotes the normal derivative
and $\partial  {\cal R}$ is the surface delimiting the region ${\cal R}$.

The Dirichlet $G_D$ and Neumann $G_N$ Green's function satisfy the
boundary conditions
\begin{equation}
\label{eq:def-Neumann-cond}
G_{D}(\r,\r')=0 \qquad \mathrm{and}\qquad
\partial_{n'}\, G_N(\r,\r')=-1/R^{\nu-1}
\qquad\text{for\ }\r'\in\partial{\cal R}.
\end{equation}
For Dirichlet boundary conditions, we therefore have  
\begin{equation}
\label{eq:pot-Dirichlet}
\psi(\r)=\int_{\cal R} \rho(\r')\, G_D(\r,\r')\,d\r'
-\frac{1}{S_{\nu}} \int_{\partial{\cal R}} \psi(\r')\,\frac{\partial
G_D}{\partial n'}(\r,\r')\,dS'
\end{equation}
For the isotropic situation considered here and from Gauss law,
the last term in equation~(\ref{eq:pot-Dirichlet}) may be written as
\begin{equation}
-\psi(R) \frac{1}{S_{\nu}} \int_{\partial{\cal R}}\,\frac{\partial
G_D}{\partial n'}(\r,\r')\,dS'=\psi(R).
\end{equation}
So far, the choice $G_D(\r,\r')=0$ for $\r'\in\partial{\cal R}$
does not ensure that $\psi(R)=0$. 
We chose to fix
the zero of the electric potential at the boundary of the cell
$\psi(R)=0$ so that we recover Eq. (\ref{eq:convol}).

For Neumann boundary conditions, 
\begin{equation}
\label{eq:pot-Neumann}
\psi(\r)= \left<\psi\right>_S+ \int_{\cal R} \rho(\r')
\,G_N(\r,\r')\,d\r' +\frac{1}{S_{\nu}} \int_{\partial{\cal R}}
\frac{\partial \psi(\r')}{\partial n'}\,G_N(\r,\r')\,dS',
\end{equation}
where the first term
$\left<\psi\right>_S=\psi(R)$ is the average of the electric potential
on the surface of the cell and we chose it to vanish. 
The last term in equation~(\ref{eq:pot-Neumann}) is
\begin{equation}
\label{eq:terme-qui-peut-etre-nul}
\frac{1}{S_{\nu}}\partial_n\psi(R) \int_{\partial{\cal R}}
G_N(\r,\r')\,dS'
=
-\frac{1}{R^{\nu-1}}\int_{\cal R} \rho(\r')\,d\r' \int_{\partial{\cal R}}
G_N(\r,\r'')\,dS''.
\end{equation}
We have used Gauss law: $-\partial_n
\psi(R)=\int\rho(\r')\,d\r'/R^{\nu-1}$. A proper choice of the Green's function
ensures that
the surface integral on the right-hand-side of 
(\ref{eq:terme-qui-peut-etre-nul})
is independent of $\r$, and vanishes (one can shift $G_N$
by an arbitrary constant). Explicitly, in
the three dimensional case, the choice 
\begin{eqnarray}
\label{eq:Green-N}
G_N(\r,\r')&=&-\frac{1}{R}+\frac{1}{r_{>}}+
\sum_{\ell=1}^{\infty} \left[\frac{r_{<}^{\ell}}{r_{>}^{\ell+1}}
+\frac{\ell+1}{\ell}\frac{(rr')^{\ell}}{R^{2\ell+1}}
\right]
P_{\ell}(\cos\theta)
\end{eqnarray}
with $r_{>}=\max(r,r')$ and $r_{<}=\min(r,r')$, $\theta$ the angle
between $\r$ and $\r'$ and $P_{\ell}$ the Legendre polynomial of order
$\ell$, makes the term~(\ref{eq:terme-qui-peut-etre-nul}) to vanish. 
Finally, with this choice for the Green's function and for the 
reference potential, we have in both cases of boundary conditions
\begin{equation}
\label{eq:pot-int}
\psi(\r)=
\int_{\cal R} \rho(\r')\,G(\r,\r')\,d\r' \quad \hbox{and}\quad \psi(R) = 0.
\end{equation}
A useful property that follows from these considerations 
is that for \textit{any} isotropic charge distribution
$\rho(r)$ (eventually non-globally neutral) we have in both cases of
boundary conditions
\begin{equation}
\label{eq:propriete-utile}
\int_{\cal R}
\rho(r')\,G(\mathbf{R},\r')\,d\r'= 0  
\end{equation}
where $\mathbf{R}\in\partial{\cal R}$. This follows directly in the
Dirichlet boundary conditions case from $G(\mathbf{R},\r')=0$ and in
the Neumann boundary conditions case form the particular
choice~(\ref{eq:Green-N}). Finally, we emphasize that the symmetry
property $G(\r,\r')=G(\r',\r)$ is not necessarily fulfilled by a 
generic Green's function \cite{Kim}, but may be imposed as a separate
requirement, and holds for the functions considered here. 
\end{appendix}

%%%%%%%%%%%%%%%%%%%%%%%%%%%%%%%%%%%%%%%%%%%%%%%%%%%%%%%%%%%%%%
\section*{Acknowledgments}

The visit of G.~T.~at LPT Orsay was supported by ECOS
Nord/COLCIENCIAS-ICETEX-ICFES action C00P02 of French and Colombian
cooperation. G.~T.~acknowledge partial financial support from
COLCIENCIAS and BID through project \#1204-05-10078.

%%%%%%%%%%%%%%%%%%%%%%%%%%%%%%%%%%%%%%%%%%%%%%%%%%%%%%%%%%%%%%%%%%%%%

\end{document}